\documentclass[preprint,prd,nofootinbib,tightenlines,amsmath,amssymb]{revtex4}
\usepackage{graphicx}% Include figure files
\usepackage{bm}% bold math
\usepackage{color}

\begin{document}
\baselineskip=15pt \parskip=3pt

\vspace*{3em}

\title{Low-Mass Dark-Matter Hint from CDMS II, \\ Higgs Boson at the LHC, and Darkon Models}

\author{Xiao-Gang He$^{1,2,3}$}
\author{Jusak Tandean$^{2}$}
\affiliation{$^1$INPAC, SKLPPC, and Department of Physics,
Shanghai Jiao Tong University, Shanghai, China \vspace*{1ex}
\\
$^2$Department of Physics and Center for Theoretical Sciences,
National Taiwan University, Taipei 106, Taiwan \vspace*{1ex}
\\
\mbox{$^3$Department of Physics, National Tsing Hua University,} and
National Center for Theoretical Sciences, Hsinchu 300, Taiwan \vspace*{1ex}
\\
}

%\date{\today}
%\pacs{14.80.Bn, 95.35.+d, 14.80.Ec, 12.60.-i}

\begin{abstract}

The \mbox{CDMS II} experiment has observed three events which may have arisen from
weakly interacting massive particle (WIMP) dark matter (DM) with mass of order~9~GeV colliding
with nuclei.
Although the implied WIMP parameter region seems to be excluded by limits from the XENON
experiments, it is interesting that most of this tension can go away if
the WIMP-nucleon interaction violates isospin.
This motivates us to explore some of the implications for models in which a~real gauge-singlet
scalar particle, the darkon, serves as the~WIMP,
taking into account the recent discovery of a Higgs boson at the~LHC and Planck determination
of the DM relic density.
In the simplest scenario, involving only the standard model plus a darkon, the Higgs
boson is largely invisible due to its decay into a pair of darkons having the WIMP mass
suggested by \mbox{CDMS II} and hence cannot be identified with the one found at the LHC.
We find, on the other hand, that a~two-Higgs-doublet model supplemented with a darkon has ample
parameter space to accommodate well both the new potential DM hint from \mbox{CDMS II} and
the Higgs data from the LHC, whether or not the darkon-nucleon interaction conserves isospin.

\end{abstract}

\maketitle

\section{Introduction}

The latest direct search for weakly interacting massive particle (WIMP) dark matter (DM)
carried out by the CDMS Collaboration has come up with a tantalizing possible hint of
WIMP collisions with ordinary matter~\cite{cdms}.
Their analysis of data collected using the CDMS\,II silicon detectors has turned up
three events in the signal region with confidence level of about three sigmas~\cite{cdms2}.
If interpreted to be due to spin-independent WIMP-nucleon scattering, the new data
favor a WIMP mass of 8.6\,GeV and scattering cross-section of~\,$1.9\times10^{-41}{\rm\,cm}^2$.\,

Because of the relatively low statistical significance of this finding, it still does not
provide definitive evidence for the existence of WIMPs~\cite{cdms,cdms2}.
Nevertheless, it has added to the excitement previously aroused by the excess events which had
been seen in the DAMA, CoGeNT, and CRESST-II direct detection experiments and could
conceivably be of WIMP origin as well~\cite{Bernabei:2010mq,Aalseth:2012if,Angloher:2011uu}.
Like the CDMS\,II observation, the earlier findings are suggestive of a~WIMP mass in the range
roughly from 7~to~40~GeV and WIMP-nucleon scattering cross-sections of order
\,$10^{-42}$ to $10^{-40}$~cm$^2$,\, although the respective ranges preferred by the different
experiments do not fully agree with each other.

The CDMS\,II result has also contributed to the ongoing tension between these potential WIMP
indications and the null results of direct searches by the XENON Collaborations and
others~\cite{Ahmed:2009zw,Angle:2011th,Aprile:2012nq,Felizardo:2011uw,
Armengaud:2012pfa,Li:2013fla,Agnese:2013cvt,Brown:2011dp}.
This puzzle on the experimental side is yet to be resolved comprehensively, and presently
for WIMP masses under 15\,GeV the null results are still controversial~\cite{Hooper:2010uy}.
It is intriguing, however, that for most of the WIMP mass region implied by CDMS\,II
the conflict with the exclusion limits set by the latter experiments can disappear if the WIMP
interactions with nucleons are allowed to violate isospin symmetry, as will be shown later,
which is not the case with the signal regions favored by DAMA, CoGeNT, and CRESST-II.

It is then of interest to see how simple models may account for these developments
regarding the light-WIMP hypothesis, taking into account the recent discovery of a Higgs boson
with mass around 125\,GeV at the LHC~\cite{higgs@lhc} and determination of the DM relic
density by the Planck Collaboration~\cite{planck}.
In this paper, we will focus on a scenario in which a real gauge-singlet scalar particle
called the darkon plays the role of WIMP DM.
The intimate interplay between the DM and Higgs sectors of darkon models makes them appealing
to study in light of new experimental information that is relevant to either sector.

In the minimal darkon model~\cite{Silveira:1985rk,He:2010nt}, which is the standard
model (SM) slightly expanded with the addition of a darkon (SM+D), the Higgs boson having
a mass of 125\,GeV will decay predominantly into a darkon pair if the darkon mass
\,$m_D^{}\sim10$\,GeV\, as suggested by~CDMS\,II.
This Higgs boson would then be largely invisible, very unlike the one found at
the~LHC~\cite{higgs@lhc}.

In order to accommodate both a Higgs boson consistent with LHC data and a light darkon,
the~SM+D must therefore be enlarged.
One of the simplest extensions contains an extra Higgs
doublet~\cite{Bird:2006jd,Cai:2011kb,He:2011gc}.
In a two-Higgs-doublet model plus a darkon (THDM+D), the lighter $CP$-even Higgs boson can be
arranged to be SM-like and the heavier one primarily responsible for the light-darkon
annihilation which reproduces the observed DM relic density~\cite{Cai:2011kb,He:2011gc}.

In the following section, we explore some of the implications of the latest CDMS II, LHC,
and Planck measurements for the THDM+D with a low-mass darkon.
We will consider both the cases of WIMP-nucleon interactions that conserve and violate
isospin symmetry.

\section{Two-Higgs-doublet model plus darkon\label{thdm+d}}

For the Higgs sector of the model, we adopt the so-called two-Higgs-doublet model~(THDM)
of type III, in which the quarks and charged leptons couple to both Higgs doublets.
In the type-I THDM only one of the doublets couples to fermions~\cite{thdm}, and so adding
a darkon would only lead to darkon-Higgs interactions similar to those in the~SM+D.
Although the type-II THDM+D can also provide a SM-like Higgs boson and a low-mass
darkon~\cite{Cai:2011kb}, only the type-III THDM+D can offer WIMP-nucleon effective
couplings with sufficiently sizable isospin violation~\cite{He:2011gc}.

Its Yukawa Lagrangian has the form~\cite{thdm}
\begin{eqnarray} \label{LY}
\textrm{L}_{\rm Y}^{} \,\,=\,\,
-\bar Q_{j,L}^{}\bigl(\lambda_a^{\cal U}\bigr)_{jl}\tilde H_{a\,}^{}{\cal U}_{l,R}^{}
- \bar Q_{j,L}^{}\bigl(\lambda_a^{\cal D}\bigr)_{jl} H_a^{} {\cal D}_{l,R}^{}
- \bar L_{j,L}^{}\bigl(\lambda_a^{E}\bigr)_{jl} H_a^{}E_{l,R}^{}
\;+\; {\rm H.c.} ~,
\end{eqnarray}
where summation over \,$j,l=1,2,3$\, and \,$a=1,2$\, is implicit,
$Q_{j,L}$ $(L_{l,L})$ represent the left-handed quark (lepton) doublets,
${\cal U}_{l,R}^{}$ and ${\cal D}_{l,R}$ $(E_{l,R})$ are
the right-handed quark (charged lepton) fields, $H_{1,2}^{}$~denote the Higgs doublets,
\,$\tilde H_{1,2}^{}=i\tau_2^{}H_{1,2}^*$,\, and so $\lambda_{1,2}^{{\cal U,D},E}$ are
3$\times$3 matrices for the Yukawa couplings.
In the Higgs sector, the $CP$-even components of the doublets mix with mixing angle~$\alpha$,
while the $CP$-odd components mix, as do the charged ones, with mixing angle $\beta$.
The latter is related to the vacuum expectation values $v_{1,2}^{}$ of $H_{1,2}^{}$,
respectively, by \,$\cos\beta=v_1^{}/v$\, and \,$\sin\beta=v_2^{}/v$,\, with \,$v_1^2+v_2^2=v^2$\,
and~\,$v\simeq246$\,GeV.\,
We have followed the notation of Ref.\,\cite{He:2011gc} which has a more detailed description
of the model.

After the diagonalization of the fermion mass matrices, the flavor-diagonal couplings of
the physical $CP$-even Higgs fields \,${\cal H}=h,H$\, to the fermion mass eigenstate $f$
can be described by
\begin{eqnarray}
{\cal L}_{ff\cal H}^{} \,\,=\,\, -k_f^{\cal H}\,m_f^{}\,\bar f f\,\frac{\cal H}{v} ~,
\end{eqnarray}
where $m_f^{}$ is the mass of $f$ and for, say, the first family
\begin{eqnarray} & \displaystyle
k_u^h \,\,=\,\, \frac{\cos\alpha}{\sin\beta} \,-\,
\frac{\lambda_1^u\,v\,\cos(\alpha-\beta)}{\sqrt2\,m_u^{}\,\sin\beta} ~, \hspace{5ex}
k_u^H \,\,=\,\, \frac{\sin\alpha}{\sin\beta} \,-\,
\frac{\lambda_1^u\,v\,\sin(\alpha-\beta)}{\sqrt2\,m_u^{}\,\sin\beta} ~,
& \nonumber \\ & \displaystyle
k_{d,e}^h \,\,=\,\, -\frac{\sin\alpha}{\cos\beta} \,+\,
\frac{\lambda_2^{d,e}v\,\cos(\alpha-\beta)}{\sqrt2\,m_{d,e}^{}\,\cos\beta} ~, \hspace{5ex}
k_{d,e}^H \,\,=\,\, \frac{\cos\alpha}{\cos\beta} \,+\,
\frac{\lambda_2^{d,e}v\,\sin(\alpha-\beta)}{\sqrt2\,m_{d,e}^{}\,\cos\beta} ~, \label{kf}
\end{eqnarray}
with \,$\lambda_{a}^{u,d,e}=\bigl(\lambda_{a}^{{\cal U,D},E}\bigr){}_{11}^{}$.\,
The corresponding $k_f^{\cal H}$ for the other two families have analogous expressions.
Since only \,$\lambda_1^f v_1^{}+\lambda_2^f v_2^{}=\sqrt2\,m_f^{}$\, is fixed
by the $f$ mass, $\lambda_{a}^f$~in $k_f^{\cal H}$ is a free parameter, and so is~$k_f^{\cal H}$.
Setting \,$\lambda_1^{\cal U}=\lambda_2^{\cal D}=\lambda_2^E=0$\, would lead to the type-II
THDM+D considered in~Ref.\,\cite{Cai:2011kb}.
Since the type-III THDM is known to have flavor-changing neutral Higgs couplings at
tree level, we assume that in the THDM+D they have their naturally small values according to
the Cheng-Sher {\it ansatz}~\cite{Cheng:1987rs}, namely
\,$(\lambda_a)_{jl}^{}\sim \bigl(m_j^{}m_l^{}\bigr){}^{1/2}/v$\, for~\,$j\neq l$.\,
If necessary, the effects of these flavor-changing couplings could be further suppressed by
increasing the mediating Higgs masses.

In the DM sector of the THDM+D, the stability of the darkon, $D$, as a WIMP candidate is ensured
by requiring it to be a gauge singlet and imposing a~discrete $Z_2$ symmetry under which
$D$ is odd and all the other fields are even.
The renormalizable Lagrangian for $D$ is then~\cite{Bird:2006jd}
\begin{eqnarray} \label{LD2hdmd}
{\cal L}_D^{} \,\,=\,\, \mbox{$\frac{1}{2}$}\partial^\mu D\,\partial_\mu^{}D
-\mbox{$\frac{1}{4}$}\lambda_D^{}D^4 - \mbox{$\frac{1}{2}$}m_0^2D^2 \,-\,
\bigl[ \lambda_1^{}H_1^{\dag}H_1^{} + \lambda_2^{}H^\dagger_2 H_2^{} +
\lambda_3^{}\bigl(H_1^{\dag}H_2^{}+H_2^\dagger H_1^{}\bigr) \bigr] D^2 ~.
\end{eqnarray}
After electroweak symmetry breaking, ${\cal L}_D^{}$ includes the darkon mass
$m_D^{}$ and the $DD(h,H)$ terms \,$-\lambda_h^{}v\, D^2 h-\lambda_H^{}v\,D^2 H$\, with
\begin{eqnarray} & \displaystyle
m_D^2 \,\,=\,\, m_0^2 \,+\,
\bigl[\lambda_1^{}\cos^2\!\beta+\lambda_2^{}\sin^2\!\beta+\lambda_3^{}\sin(2\beta)\bigr]v^2 ~,
& \nonumber \\ & \displaystyle
\lambda_h^{} \,\,=\,\, -\lambda_1^{}\sin\alpha\,\cos\beta+\lambda_2^{}\cos\alpha\,\sin\beta +
\lambda_3^{}\cos(\alpha+\beta) ~, & \nonumber \\ &
\lambda_H^{} \,\,=\,\,
\lambda_1^{}\cos\alpha\,\cos\beta+\lambda_2^{}\sin\alpha\,\sin\beta +
\lambda_3^{}\sin(\alpha+\beta) ~, & \label{lambda}
\end{eqnarray}
but no $DDA$ coupling involving the physical $CP$-odd Higgs boson $A$ if $CP$ is conserved.
Since $m_0^{}$ and $\lambda_{1,2,3}^{}$ are free parameters, so are $m_D^{}$
and~$\lambda_{h,H}^{}$.

To evaluate the darkon annihilation rates, the couplings of $h$ and $H$ to the $W$ and $Z$ bosons
may also be pertinent depending on~$m_D^{}$.
They are given by~\cite{thdm}
\begin{eqnarray} \label{vvh}
{\cal L}_{VV\cal H}^{} \,\,=\,\, \frac{1}{v}
\bigl(2m_W^2\,W^{+\mu}W_\mu^-+m_Z^2\,Z^\mu Z_\mu^{}\bigr)
\bigl[h\,\sin(\beta-\alpha)+H\, \cos(\beta-\alpha)\bigr]
\end{eqnarray}
from the Higgs kinetic sector of the model.

We start our numerical work by identifying the lighter Higgs particle $h$ with the Higgs
boson observed at the LHC and fixing the couplings discussed above.
The latest measurements on $h$ have begun to indicate that the particle has SM-like
properties, but there is still some room in its couplings for deviations from SM expectations.
Specifically, according to a number of analyses~\cite{h2inv}, the current data imply that
the $h$ couplings to the $W$ and $Z$ bosons cannot differ from their SM values by more
than~${\cal O}(10\%)$, whereas the couplings to fermions are less well-determined.
Furthermore, the branching ratio of nonstandard decays of $h$ into invisible or undetected
final-states can be as high as a few tens percent~\cite{h2inv}.
All this implies that in general the free parameters $k_f^h$ and $\sin(\beta-\alpha)$ may
deviate from unity accordingly and that for a light darkon $\lambda_h^{}$ can be nonzero,
but not large.
For definiteness and simplicity, we take \,$\beta-\alpha=\pi/2$\, and \,$\lambda_h=0$,\,
following Ref.\,\cite{He:2011gc}.
Consequently, at tree level $h$ has fermionic and gauge couplings identical to their SM
counterparts, in particular~\,$k_f^h=1$,\, but no interaction with the darkon, preventing
$h$ from having a nonnegligible invisible decay mode.
Moreover, the heavier Higgs boson $H$ now couples to fermions and the darkon according to
\begin{eqnarray} \label{kH}  & \displaystyle
k_u^H \,\,=\,\, -\cot\beta\,+\,\frac{\lambda_1^u\,v}{\sqrt2\,m_u^{}\,\sin\beta} ~, \hspace{5ex}
k_{d,e}^H \,\,=\,\, \tan\beta \,-\, \frac{\lambda_2^{d,e}\,v}{\sqrt2\,m_{d,e}^{}\cos\beta} ~, &
\\ \label{lambdahH} & \displaystyle
\lambda_H^{} \,\,=\,\, \mbox{$\frac{1}{2}$}\bigl(\lambda_1^{}-\lambda_2^{}\bigr)\sin(2\beta)
- \lambda_3^{}\cos(2\beta) ~, ~~~~~~~ &
\end{eqnarray}
but no longer has tree-level couplings to $W$ and $Z$, the $k_f^H$ formulas for the second
and third families being analogous.
It follows from these choices that for a light darkon, with \,$m_D^{}<m_h^{}/2$,\,
its annihilation occurs mainly via $H$-mediated diagrams.

The darkon being the DM candidate, its annihilation cross-section must reproduce the observed
DM relic density~$\Omega$.
Its most recent value has been determined by the Planck Collaboration from the Planck
measurement and other data to be \,$\Omega\bar h^2=0.1187\pm0.0017$\,~\cite{planck},
where $\bar h$ is the Hubble parameter.
To extract $\lambda_H^{}$ for specific $m_D^{}$ and $H$-mass, $m_H^{}$, values, after
$k_f^H$ are chosen, we require the darkon relic density to satisfy the 90\% CL (confidence level)
range of its experimental value,~\,$0.1159\le\Omega\bar h^2\le0.1215$.\,
We can then employ the obtained $\lambda_H^{}$ to predict the darkon-nucleon scattering
cross-section and compare it with the direct search data.
We will treat in turn the cases where the darkon-nucleon interactions respect and violate
isospin symmetry.

\begin{figure}[t]
\includegraphics[width=92mm]{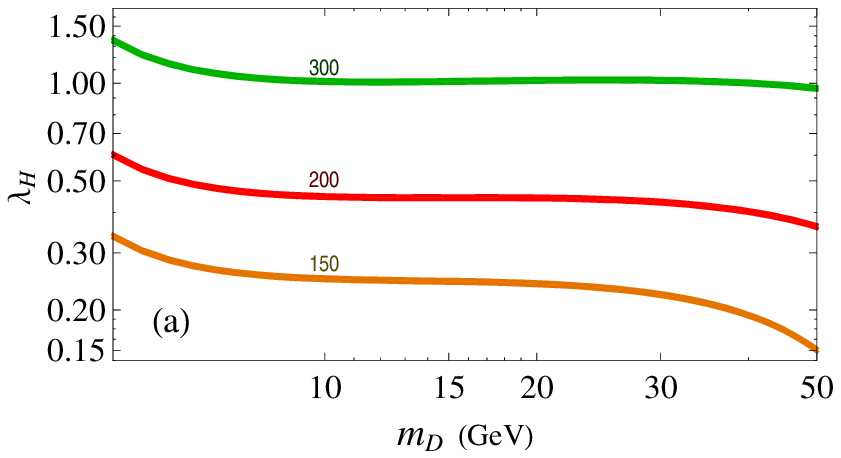}\vspace{-1ex}\\
\includegraphics[width=97mm]{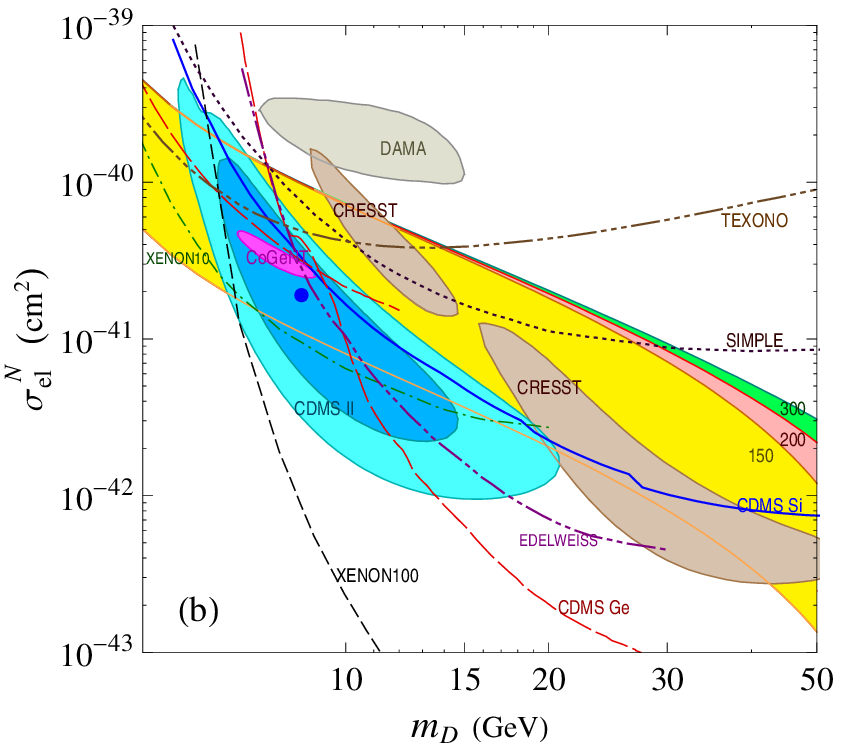} \, \,\vspace{-1ex}
\caption{(a)~Darkon-$H$ coupling $\lambda_H^{}$ as a function of darkon mass $m_D^{}$ for
\,$m_H^{}=150,200,300$~GeV,\, with the other couplings specified in the text, in the THDM+D
with isospin-conserving darkon-nucleon interactions.
(b)~The resulting darkon-nucleon scattering cross-section~$\sigma_{\rm el}^{N}$, compared~to
90\%\,CL~upper limits from XENON10 (green dashed-dotted curve)~\cite{Angle:2011th},
XENON100 (black short-dashed curve)~\cite{Aprile:2012nq},
CDMS~Ge (red long-dashed curves)~\cite{Ahmed:2009zw},
CDMS~Si (blue solid curve)~\cite{Agnese:2013cvt},
Stage 2 of SIMPLE (dark dotted curve)~\cite{Felizardo:2011uw},
EDELWEISS  (purple dashed-double-dotted curve)~\cite{Armengaud:2012pfa}, and
TEXONO (brown dashed-triple-dotted curve)~\cite{Li:2013fla}.
Also plotted are a~gray patch compatible with the DAMA Na modulation signal at the 3$\sigma$
level~\cite{Savage:2008er},
two \mbox{2$\sigma$-confidence} (light brown) areas representing the CRESST-II
result~\cite{Angloher:2011uu},
the 90\%\,CL (magenta) signal region suggested by CoGeNT~\cite{Aalseth:2012if}, and
a blue (cyan) area for a possible signal at 68\% (90\%) CL from CDMS\,II~\cite{cdms},
with the blue dot marking the maximum likelihood point at
$\bigl(8.6{\rm\,GeV},1.9\times10^{-41}{\rm\,cm}^2\bigr)$.\label{lambda_2hdm+d}}
\end{figure}

In the first case, since $\lambda_1^u$ and $\lambda_2^{d,e}$ in Eq.\,(\ref{kH}) as well as 
the corresponding parameters for the other two families are free parameters, 
again following Ref.\,\cite{He:2011gc} we pick  for definiteness \,$k_f^H=1$.\,
We present the $\lambda_H^{}$ ranges allowed by the relic data for the low-mass region
\,$5{\rm\,GeV}\le m_D^{}\le50$\,GeV\, and some illustrative values
of $m_H^{}$ in~Fig.\,\ref{lambda_2hdm+d}(a),
where the width of each band reflects the 90\%\,CL range of~$\Omega\bar h^2$\,
above.\footnote{\baselineskip=13pt%
With the simple choices \,$k_f^H=1$,\, the rate of darkon annihilation into $b\bar b$ seems
to be in somewhat of a tension with upper limits inferred from searches for DM signals in
diffuse gamma-ray data from the Fermi Large Area Telescope observations of dwarf
spheroidal satellite galaxies of the Milky Way~\cite{Ackermann:2011wa}.
This can be circumvented by taking instead \,$k_b^H<1$\, and also adjusting
the other $k_f^H$ to satisfy any additional constraints.\smallskip}

The darkon-nucleon elastic scattering occurs via an $H$-mediated diagram in the $t$ channel.
Its cross section is given by~\cite{He:2011gc}
\,$\sigma_{\rm el}^N=
\lambda_H^2 g_{NNH}^2 v^2 m_N^2/\bigl[\pi\bigl(m_D^{}+m_N^{}\bigr){}^2m_H^4\bigr]$,\,
where $m_N^{}$ is the average of the proton and neutron masses.
The effective $H$-nucleon coupling $g_{NNH}$, which respects isospin, has a rather wide
range,~\,$0.0011\le g_{NNH}^{}\le0.0032$~\cite{He:2010nt,He:2011gc}, because of its dependence
on the pion-nucleon sigma term which is not well determined~\cite{Ellis:2008hf}.
We display in Fig.\,\ref{lambda_2hdm+d}(b) the calculated $\sigma_{\rm el}^{N}$ corresponding
to the parameter selections in~Fig.\,\ref{lambda_2hdm+d}(a).
The width of each predicted $\sigma_{\rm el}^{N}$ curve arises mainly from the sizable
uncertainty of~$g_{NNH}^{}$.
Also on display are the contours and curves reproduced from
Refs.~\cite{Bernabei:2010mq,Aalseth:2012if,Angloher:2011uu,Ahmed:2009zw,Angle:2011th,
Aprile:2012nq,Felizardo:2011uw,Armengaud:2012pfa,Li:2013fla,Agnese:2013cvt,Savage:2008er}
and representing the results of CDMS\,II and other latest DM direct detection experiments.
One can see that the THDM+D prediction overlaps significantly with the 68\%\,CL (90\%\,CL)
possible signal [blue (cyan)] region from CDMS\,II~\cite{cdms}, as well as with the signal
regions preferred by CoGeNT and~CRESST-II.
At the same time, the prediction and the potential signal regions mostly appear to be in
serious conflict with the exclusion limit from XENON100 and in partial disagreement with
some of the other limits.

One of the important proposals in the literature to resolve the light-WIMP inconsistencies
among the direct search data is to allow substantial violation of isospin symmetry in
the WIMP-nucleon interactions~\cite{Kurylov:2003ra,Feng:2011vu}.
It turns out that the tension can be partially alleviated if the effective WIMP couplings
$f_p^{}$ and $f_n^{}$ to the proton and neutron, respectively, obey the
ratio~\,$f_n^{}/f_p^{}\simeq-0.7$\,~\cite{Kurylov:2003ra,Feng:2011vu}.

\begin{figure}[t]
\includegraphics[width=97mm]{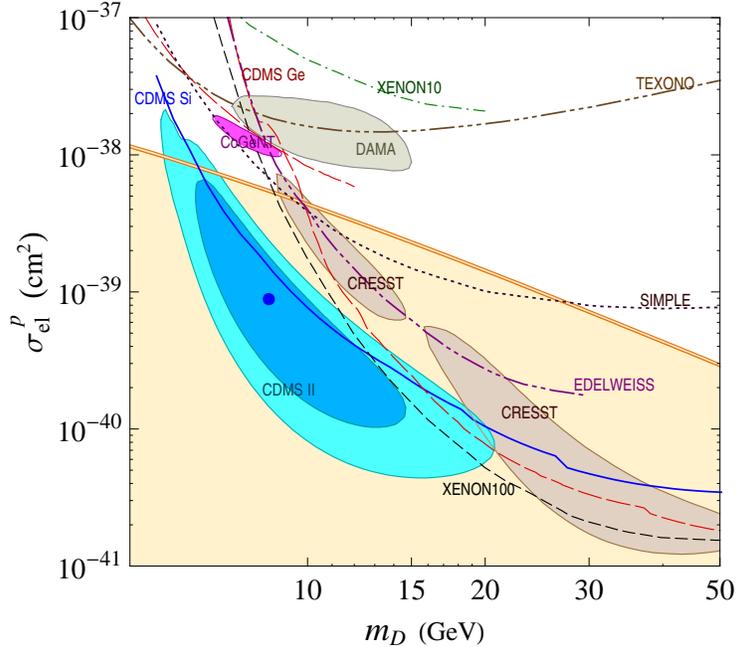}
\caption{Darkon-proton scattering cross-section $\sigma_{\rm el}^p$ in THDM+D with
isospin-violating darkon-nucleon couplings
[orange curve and (lightly shaded) light-orange region below it]
compared with the direct search results in Fig.\,\ref{lambda_2hdm+d}(b) reproduced
using WIMP-nucleon effective couplings satisfying \,$f_n^{}=-0.7f_p^{}$.\label{cs-ivdm}}
\end{figure}

We now apply this requirement to the THDM+D and the new CDMS\,II data.
Since $k_f^H$, as in~Eq.\,(\ref{kH}), are free parameters, letting them deviate from
the choices \,$k_f^H=1$\, made above, which conserve isospin, can lead to large isospin
violation in the effective interactions of the darkon with the proton and neutron.
This is achievable through the couplings of $H$ to the proton and neutron, denoted by
$g_{ppH}^{}$ and $g_{nnH}^{}$, respectively, which are related to the $k_f^H$  for
quarks by~\,$g_{{\cal NN}H}^{}=\mbox{\small$\sum$}_{q\,}g_q^{\cal N}k_q^H$,\,
where \,${\cal N}=p,n$,\, the sum is over all quarks, and $g_q^{\cal N}$ results from
the $\cal N$ matrix-element of the $q$ scalar-density~\cite{He:2011gc,Ellis:2008hf}.
Thus, one can arrive at substantial isospin violation with \,$k_u^H\neq k_d^H$\, and
the other $k_q^H$ being sufficiently small.
This has been done in Ref.\,\cite{He:2011gc}, with $g_{nnH}^{}/g_{ppH}^{}$ fixed to
the desired value of $f_n^{}/f_p^{}$.
The result is that, although the prediction is too low compared to the DAMA and CoGeNT
preferred areas by a factor of a few, sizable part of it escapes the stringent bound from
XENON100 as well as the new limit from CDMS\,II silicon detector data~\cite{Agnese:2013cvt}.
All this is illustrated in~Fig.\,\ref{cs-ivdm}, where the calculated darkon-proton scattering
cross-section, $\sigma_{\rm el}^p$, is compared to the contours and curves translated
from the direct search data shown in Fig.\,\ref{lambda_2hdm+d}(b) using the expression provided
in Ref.\,\cite{Feng:2011vu} which relates the WIMP-nucleon and WIMP-proton cross-sections,
with \,$f_n^{}=-0.7f_p^{}$\, imposed.
The orange curve in Fig.\,\ref{cs-ivdm} indicates the maximum prediction, corresponding to
\,$\lambda_H^{}k_u^H={\cal O}\bigl(10^3\bigr)$,\, \,$k_u^{H}\sim-2k_d^{H}$,\, and
the other $k_f^H$ being negligible by comparison~\cite{He:2011gc}.\footnote{The enhanced size
of $k_{u,d}^H$ confirms the finding of Ref.\,\cite{Gao:2011ka}.\vspace{-2ex}}
The lightly shaded (light orange) region below the orange curve corresponds to the prediction
with other choices of $k_f^H$ subject to the relic data and \,$f_n^{}=-0.7f_p^{}$\, requirements.

It is interesting to notice that in~Fig.\,\ref{cs-ivdm} nearly all of the blue (cyan) area
representing the 68\%\,CL (90\%\,CL) possible CDMS\,II signal is allowed by all of
the present limits, although it no longer overlaps with the CoGeNT (magenta) patch.
This is in stark contrast to the signal regions favored by DAMA, CoGeNT, and CRESST-II,
almost all of which are excluded by the various limits shown.
Moreover, most of the allowed regions of CDMS\,II are within the prediction range (light
orange region).
Thus the THDM+D with a light darkon is very consistent with this new WIMP inkling,
whether the WIMP-nucleon interaction conserves isospin or~not.
Data from future direct detection experiments can be expected to provide extra tests on
the light-WIMP hypothesis and therefore also probe the darkon model further.

\section{Conclusions\label{concl}}

The three excess events detected by \mbox{CDMS II} could be the first evidence of WIMP DM
collisions with ordinary matter.
Most of the WIMP parameter space implied by this data can evade all the bounds from
other direct detection experiments if the WIMP interactions violate isospin significantly.
We have explored this new development within the context of a two-Higgs-doublet model
slightly expanded with the addition of a~real gauge-singlet scalar particle, the darkon,
acting as the~WIMP, taking into account the Higgs data from the~LHC and Planck
determination of the relic density.
We find that this model can comfortably account for both the discovered Higgs
boson and the low-mass WIMP that may have been observed by CDMS II.

\acknowledgments

This work was supported in part by MOE Academic Excellence Program (Grant No. 102R891505) and
NSC of ROC and by NNSF (Grant No. 11175115) and Shanghai Science and Technology Commission
(Grant No. 11DZ2260700) of PRC.

\end{document}